\documentclass[letter]{aa} 
\usepackage{graphicx}
\usepackage{subfigure}
\usepackage{adjustbox}
\usepackage{txfonts}
\begin{document}

   \title{The detection of magnetic chemically peculiar stars using Gaia BP/RP spectra}

   \author{ E.~Paunzen\inst{1}
   \and M.~Pri{\v s}egen\inst{2,1}
   }

   \institute{Department of Theoretical Physics and Astrophysics, Faculty of Science, Masaryk University, Kotl\'{a}\v{r}sk\'{a} 2, 611 37 Brno, Czech Republic, \email{epaunzen@physics.muni.cz}
    \and Advanced Technologies Research Institute, Faculty of Materials Science and Technology in Trnava, Slovak University of Technology in Bratislava, Bottova 25, 917 24 Trnava, Slovakia}

\date{}
 
  \abstract
   {The magnetic chemically peculiar (mCP) stars of the upper main sequence 
   are perfectly suited to studying the effects of rotation, diffusion, mass-loss, accretion,
   and pulsation in the presence of an organized stellar magnetic field. Therefore, many
   important models can only be tested with this star group.}
   {In this case study we investigate the possibility of detecting the characteristic 
   520\,nm flux depression of mCP stars using low-resolution BP/RP spectra of the 
   $Gaia$ mission. This would enable us to effectively search for these objects
   in the ever-increasing database.}
   {We employed the tool of $\Delta$a photometry to trace the 520\,nm flux depression
   for 1240 known mCP and 387 normal-type objects including binaries. To this end, we 
   folded the filter curves with the BP/RP spectra and generated the well-established
   color-color diagram.}
   {It is clearly possible to distinguish mCP stars from normal-type objects. The 
   detection rate is almost 95\% for B- and A-type objects. It then drops 
   for cooler-type  stars, which is in line with models of the 520\,nm flux depression.}
   {The BP/RP spectra are clearly qualified to efficiently search for and detect mCP stars.}


   \maketitle
%

\section{Introduction} \label{introduction} 

The chemically peculiar (CP) stars with 
spectral types from early B to early F are traditionally characterized by the 
presence of certain stellar absorption lines of abnormal strength or weakness that indicate 
peculiar surface abundances \citep{1974ARA&A..12..257P}. 
Current theories ascribe the observed chemical peculiarities to the interplay 
between gravitational settling (atomic diffusion) and 
radiative levitation \citep{2000ApJ...529..338R}.
Furthermore, convection, He settling, mass loss, and turbulent mixing can significantly
affect the chemical composition of CP stars \citep{2005A&A...443..627T}.

Many CP stars possess stable and globally organized magnetic fields with strengths of 
up to several tens of kG \citep{2008AstBu..63..139R}. These stars are often referred 
to as magnetic chemically peculiar (mCP) stars in the literature.
The origin of the magnetic field is still a matter of some 
controversy \citep{2004Natur.431..819B}. 
Recently, there was also classical pulsation among mCP stars found 
\citep{2021mobs.confE..27H}.
Therefore, these objects are perfect astrophysical laboratories for investigating the most
important phenomena occurring on the upper main sequence in the presence of a stellar
magnetic field. 

\citet{2020A&A...640A..40H} made use of the 520\,nm flux depression, typically for 
mCP stars, to search for these objects by classifying spectra from the 
Large Sky Area Multi-Object Fiber Spectroscopic 
Telescope \citep[LAMOST,][]{2012RAA....12.1197C}. It
was shown \citep{2007A&A...469.1083K} that Fe is the principal contributor to 
the 520\,nm depression for the whole range of  effective temperatures
($T_{\rm eff}$) of mCP stars, 
while Cr and Si play a role primarily in the low $T_{\rm eff}$ region.
However, a strong 520\,nm flux depression was not found for all mCP stars
\citep{2006MNRAS.372.1804K}. The reason for this is most probably the 
individual magnetic field
configuration together with the line of sight. 

Motivated by the success of \citet{2020A&A...640A..40H}, we employed the $\Delta$a
photometric system which measures the above-mentioned flux depression 
\citep{2005A&A...441..631P,2014A&A...562A..65S} to the low-resolution
blue photometer and red photometer (BP/RP) spectra of the $Gaia$ mission \citep{2021A&A...652A..86C}.
These spectra cover the wavelength region from 330 to 1050\,nm
with a resolving power between 25 and 100, depending on the
wavelength. The signal-to-noise ratio
 depends on the apparent magnitude and color of the object.
The Gaia consortium put a lot of effort into bringing all spectra 
onto a common flux and pixel (pseudo-wavelength) scale, taking 
into account variations over the focal plane. Finally, they
produced a mean spectrum from
all the observations of the same source \citep{2022arXiv220606205M}. The
spectra are available in a continuous representation, with a subset of
sources having spectra in a sampled form as well.
The $Gaia$ DR3 already includes about 219\,000\,000 mean BP/RP spectra
for objects up to the 18th magnitude in $G$. 

The spectra are therefore perfectly suited to performing a case study if the known mCP stars
can be efficiently detected using the 520\,nm flux depression and synthetic
$\Delta$a magnitudes.

  \begin{figure}
   \centering
   \includegraphics[width=0.45\textwidth]{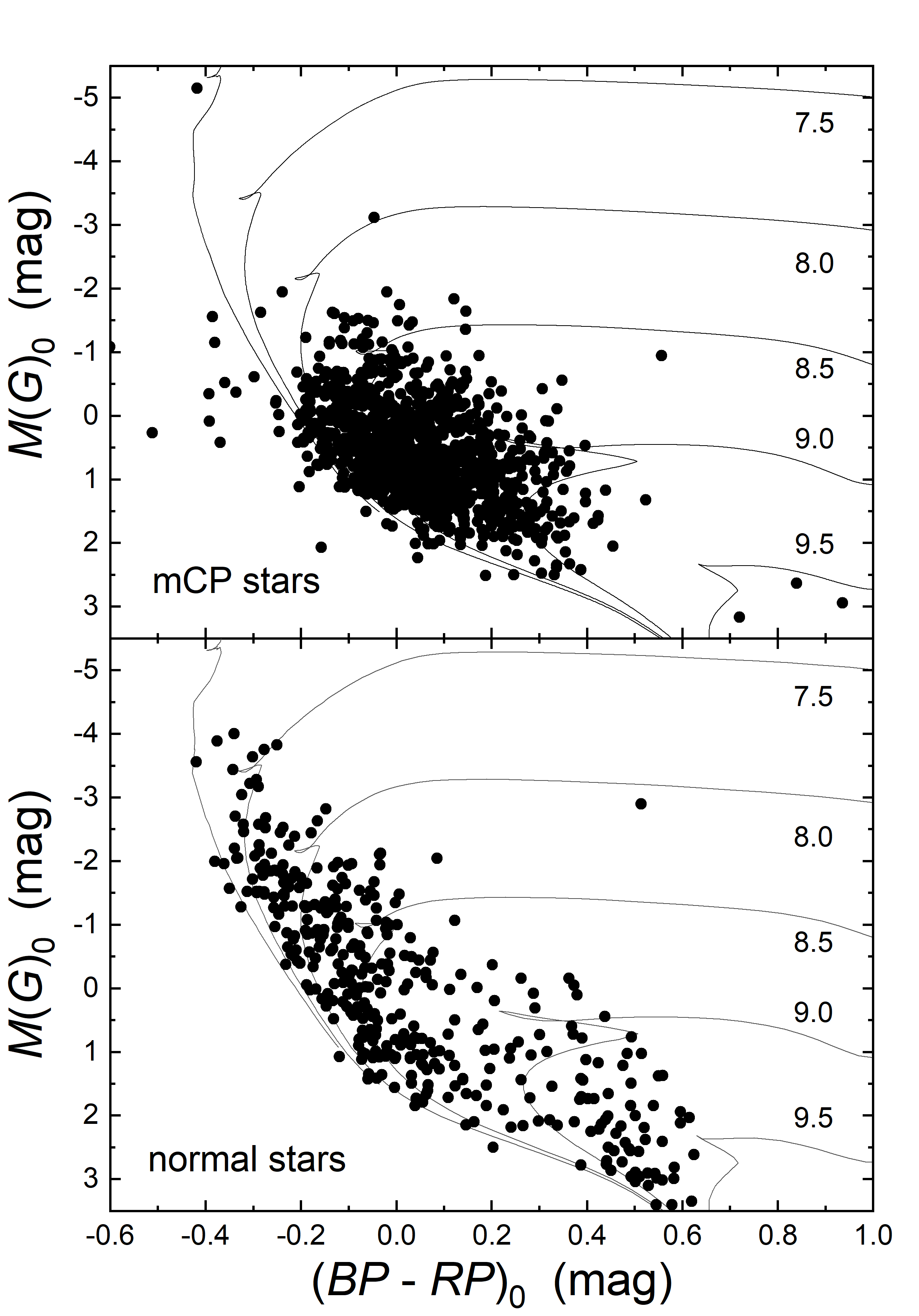}
      \caption{Diagram of  $M(G)_0$ vs $(BP-RP)_0$   for the mCP (upper panel) and
      normal-type stars (lower panel) together with isochrones from 
      \citet{2012MNRAS.427..127B} for solar metallicity [Z]\,=\,0.0152\,dex. The two samples
        overlap.} 
         \label{plot_CMD}
   \end{figure}

\section{Target selection} \label{target_selection}

For the apparent normal stars, we  chose all the objects included in the
papers by \citet{2005A&A...444..941P,2006A&A...458..293P}, who presented an
empirical $T_{\rm eff}$ calibration for B- to F-type stars. In total,
387 stars also have an available BP/RP spectrum. We note that
this sample also includes binary stars. Among them, 182 objects have a 
``Dup'' flag equal to one and/or a RUWE parameter larger than 1.5, which are both
good indicators of binarity \citep{2022AJ....163...33Z}.

The mCP stars were chosen from the following references; some stars were listed
more than once:

\begin{itemize}
    \item \citet{2005A&A...441..631P}: We
     selected all stars with $\Delta$a\,$>$\,+20\,mmag from this catalogue of $\Delta$a measurements.
    \item \citet{2019ApJ...873L...5C}: They published a list of 157 mCP stars with
    resolved magnetically split lines from the Sloan Digital Sky Survey (SDSS)/Apache 
    Point Observatory Galactic Evolution Experiment (APOGEE) survey. 
    \item \citet{2020A&A...640A..40H}: Spectra from the LAMOST Survey
    showing a 520\,nm flux depression were used to find a sample of 1002 mCP stars.
\end{itemize}

From this large selection, 1240 mCP stars have a BP/RP spectrum. For both samples,
we made no cut according to the signal-to-noise ratios of the spectra.
To check if both samples are overlapping in the color--magnitude diagram (CMD),
we used the photometry and distances from the $Gaia$ DR3 
\citep{2021AJ....161..147B,2022arXiv220800211G}. Furthermore, the reddening values
of the individual stars were taken from 
\citet{2005A&A...444..941P,2006A&A...458..293P,2008A&A...491..545N,2020A&A...640A..40H}.
If no value was found, the reddening map of \citet{2019ApJ...887...93G} was applied.
The individual reddening values were transformed to the $Gaia$ filters according
to the recommendations by the $Gaia$ 
consortium\footnote{\url{https://www.cosmos.esa.int/web/gaia/edr3-extinction-law}}.
We note that most stars are in the solar neighborhood  ($d$\,$<$\,500\,pc),
and therefore the reddening is not significant. 

In Fig. \ref{plot_CMD} the $M(G)_0$ versus $(BP-RP)_0$ diagram 
is presented for the two samples,
together with isochrones from \citet{2012MNRAS.427..127B} for solar 
metallicity [Z]\,=\,0.0152\,dex. The normal-type stars (including binaries)  cover the
whole spectral range and the main sequence very well. No stars can be found below the zero age
main sequence, which provides confidence in our method and reddening selection. 
The mCP stars, except for a few outliers, are located in the same region in the CMD (i.e.,
in the same astrophysical parameter space).
   
     \begin{figure*}
   \centering
   \includegraphics[width=0.95\textwidth]{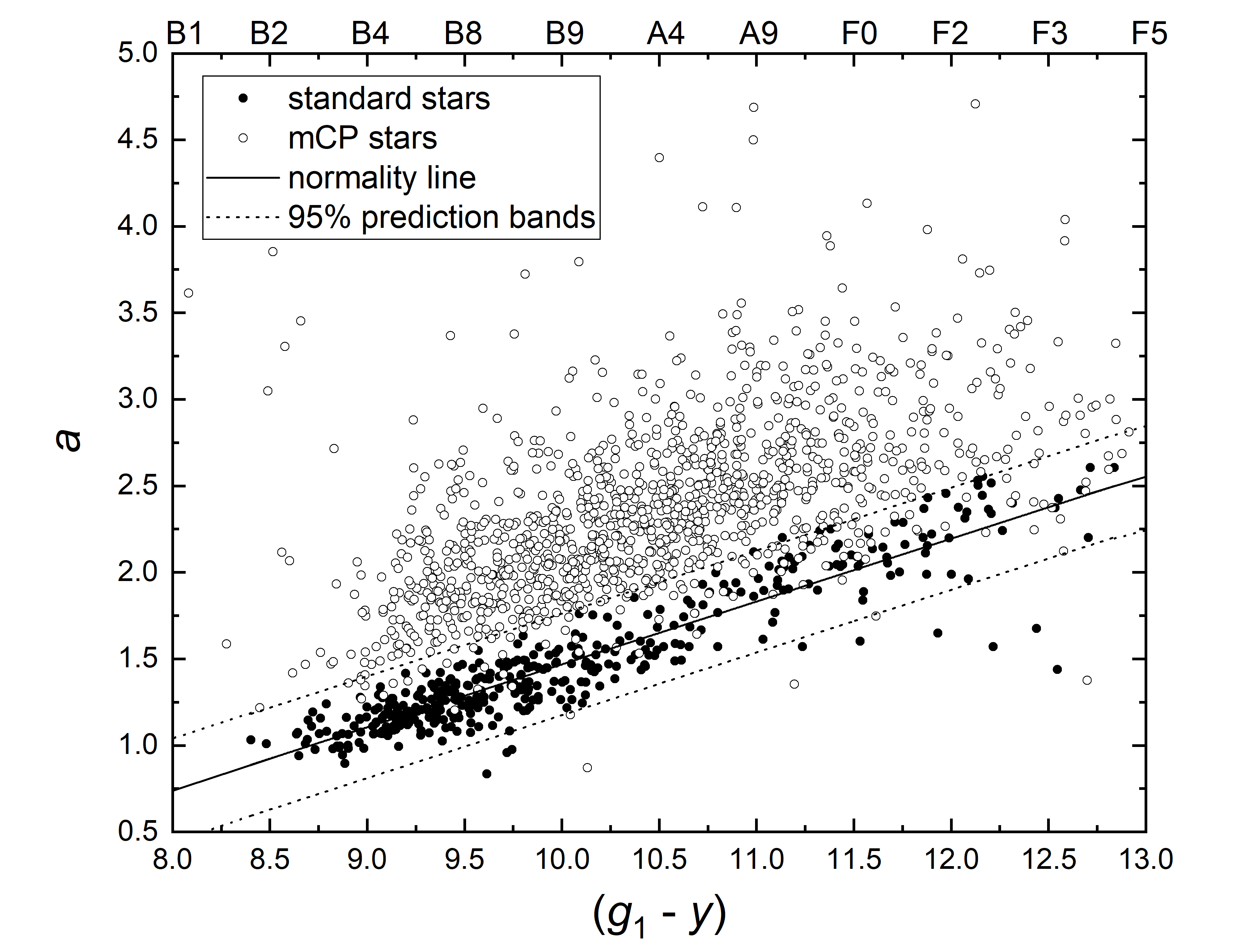}
      \caption{The $a$ versus ($g_1-y$) diagram for the 387 normal type (filled symbols) 
      and 1240 mCP
          stars (open symbols). The normality line $a = -2.16(8) + 0.363(8) (g_1-y)$
          is defined as in the classical $\Delta$a photometric system. The dotted lines are
          the 95\% prediction bands to select mCP stars. We also indicated the approximate
          spectral types according to the CMD (Figure. \ref{plot_CMD}).} 
         \label{plot_Delta_a}
   \end{figure*}
   
\begin{table}
\begin{center}
\caption{Statistics of the detection level for the synthetic $\Delta$a values
(Figure \ref{plot_Delta_a}). The numbers for ``+'' and ``$-$'' mean above and
below the 95\% prediction band of the normality line. The listed ratios 
should   ideally be 100 for the 
mCP stars and
0 for the normal-type stars.}
\label{result_classification}
\tiny
\begin{tabular}{cccc|ccc}
\hline
($g_1-y$) & $N_{\mathrm CP}-$ & $N_{\mathrm CP}$+ & \% & 
$N_{\mathrm norm}-$ & $N_{\mathrm norm}$+ & \% \\
\hline
8.0$-$8.5 & 0 & 4 & 100 & 2 & 0 & 0 \\
8.5$-$9.0 & 3 & 25 & 89 & 32 & 0 & 0 \\
9.0$-$9.5 & 7 & 132 & 95 & 106 & 0 & 0 \\
9.5$-$10.0 & 12 & 202 & 94 & 85 & 0 & 0 \\
10.0$-$10.5 & 16 & 240 & 94 & 48 & 0 & 0 \\
10.5$-$11.0 & 15 & 247 & 94 & 25 & 0 & 0 \\
11.0$-$11.5 & 23 & 150 & 87 & 29 & 1 & 3 \\
11.5$-$12.0 & 13 & 82 & 86 & 26 & 0 & 0 \\
12.0$-$12.5 & 9 & 36 & 80 & 16 & 1 & 6 \\
12.5$-$13.0 & 10 & 14 & 58 & 7 & 0 & 0 \\
\hline
\end{tabular}
\end{center}
\end{table}

\section{Analysis} \label{analysis}

To calculate the synthetic magnitudes in the $\Delta$a photometric system, we
used the approach  published in \citet{2014A&A...562A..65S}.
We first normalized all spectra to the flux given at 402\,nm.
Then they were interpolated in the wavelength region from 480 to 580\,nm
to a one-pixel resolution of 0.1\,nm applying a standard polynomial technique. 
This works perfectly because this wavelength region is very smooth.

For the filter curves,
the central wavelengths as published by \citet{1980A&A....83..328M}  were employed: $g_1$ (501\,nm),
$g_2$ (521.5\,nm), and $y$ (548.5\,nm) with a bandwidth of 13\,nm. 
According to \citet{2014A&A...562A..65S}, this results in the most significant 
$\Delta$a values.
The final magnitudes for the three filters are in arbitrary units. It is not
straightforward to determine the errors of the magnitudes. Because we did not limit 
the signal-to-noise ratios (S/N) of the spectra, the source of this error was taken into
account. For each spectrum, we took the given S/N and
randomly added an error to the individual fluxes. For the lowest S/N, values of about 10, we 
deduced an error of 1.5\%, which does not affect the conclusions and detection limits.

As the last step, we generated the $a$ versus ($g_1-y$) diagram for the normal type and mCP
stars, which is shown in Figure \ref{plot_Delta_a}. As can be seen, the normal stars
show a distinct band and are clearly separated from the mCP objects. 
We determined the normality line as in the classical $\Delta$a photometric system (i.e.,
assuming that all stars exhibit the same interstellar reddening; peculiar objects deviate 
from it by more than 3$\sigma)$. For the normal-type star sample, it was calculated as
\begin{equation}
    a = -2.16(8) + 0.363(8) (g_1-y)
\end{equation}
in the units of our synthetic photometric system. The characteristics are
similar to the observed normality 
lines for open clusters 
\citep{2007A&A...462..591N,2014A&A...564A..42P}. The color ($g_1-y$) is 
well correlated with the
$T_{\rm eff}$, especially from late B- to early F-type stars.
We split the sample into ($g_1-y$) bins and counted how many stars are 
below (``$-$'') and 
above
(``+'') the normality line. The latter is the region where the mCP stars are
located. In Table \ref{result_classification} the results are presented for both
samples. In total, from 387 normal stars, only 2 are located slightly above the 
prediction band of the region of the mCP objects.
From the 1240 mCP stars, 108 (8.7\%) are below the upper 95\% prediction band. 
However, the distribution
of the latter is homogeneous over the ($g_1-y$) range. Most of them
are located among cooler-type objects. This effect is already known, as 
\citet{2006A&A...448.1153K,2007A&A...469.1083K} showed that the capability 
of the $\Delta$a photometric system drops significantly for stars with $T_{\rm eff}$
lower than 8000\,K. This is due to the observational fact that the 520\,nm 
flux depression becomes less prominent compared to the overall line blanketing
\citep{2004MNRAS.352..863K}. For the B- and A-type stars we achieved detection 
rates of almost 95\%, which is equavalent to a photometric accuracy of 10\,mmag 
for classical CCD observations \citep{2005A&A...441..631P}.
   
\section{Conclusions} \label{conclusions}

We  showed that, with the help of the low-resolution BP/RP spectra of the
$Gaia$ mission, it is possible to efficiently search for and detect classical magnetic 
chemically peculiar  (mCP) stars of the upper main sequence. This was done by 
calculating magnitudes within the $\Delta$a photometric system which traces
the flux depression at 520\,nm typically for this star group. The latter are
benchmark objects for studying the effects of rotation, diffusion, mass loss,
and pulsation in the presence of an organized local stellar magnetic field. 

Using well-known mCP stars and a sample of normal type stars (including about 50\% binaries),
an almost 95\% detection rate for B- and A-type stars was achieved. With our
technique it is therefore possible to pre-select the best candidates for further
spectroscopic and photometric studies. 
Querying the different catalogues of the \textit{Gaia} DR3, we found spectral
tags indicating approximately 13 million B- and A-type stars up to $G$\,=\,17.65\,mag
among the published data release. \citet{2005MNRAS.362.1025P} estimated the lower 
incidence of all CP stars in the Milky Way of 6\% with a published maximum value 
of at least 16\%, respectively. The mCP stars account for about one-third of all
CP objects. Therefore, we adopt a mean value of 3\% for the 
considered spectral type range. The ratio of mCP stars that are part of binary systems
is still inconclusive \citep{2020CoSka..50..570P}. In total, we estimate 
$\sim 2 \times 10^{5}$ mCP stars to be potentially detectable for the discussed 
accuracy of the presented synthetic $\Delta$a photometry.

\begin{acknowledgements}
This work was supported by the European Regional Development Fund, 
project No. ITMS2014+: 313011W085. 
This work presents results from the European Space Agency (ESA) space mission 
Gaia. Gaia data are being processed by the Gaia Data Processing and Analysis 
Consortium (DPAC). Funding for the DPAC is provided by national institutions, 
in particular the institutions participating in the Gaia MultiLateral Agreement 
(MLA). The Gaia mission website is https://www.cosmos.esa.int/gaia. 
The Gaia archive website is https://archives.esac.esa.int/gaia. 
This research has made use of the SIMBAD database,
operated at CDS, Strasbourg, France and
of NASA's Astrophysics Data System Bibliographic Services. 
\end{acknowledgements}

\bibliographystyle{aa} 
\bibliography{main.bib}

\begin{thebibliography}{30}
\expandafter\ifx\csname natexlab\endcsname\relax\def\natexlab#1{#1}\fi

\bibitem[{{Bailer-Jones} {et~al.}(2021){Bailer-Jones}, {Rybizki}, {Fouesneau},
  {Demleitner}, \& {Andrae}}]{2021AJ....161..147B}
{Bailer-Jones}, C.~A.~L., {Rybizki}, J., {Fouesneau}, M., {Demleitner}, M., \&
  {Andrae}, R. 2021, \aj, 161, 147

\bibitem[{{Braithwaite} \& {Spruit}(2004)}]{2004Natur.431..819B}
{Braithwaite}, J. \& {Spruit}, H.~C. 2004, \nat, 431, 819

\bibitem[{{Bressan} {et~al.}(2012){Bressan}, {Marigo}, {Girardi}, {Salasnich},
  {Dal Cero}, {Rubele}, \& {Nanni}}]{2012MNRAS.427..127B}
{Bressan}, A., {Marigo}, P., {Girardi}, L., {et~al.} 2012, \mnras, 427, 127

\bibitem[{{Carrasco} {et~al.}(2021){Carrasco}, {Weiler}, {Jordi}, {Fabricius},
  {De Angeli}, {Evans}, {van Leeuwen}, {Riello}, \&
  {Montegriffo}}]{2021A&A...652A..86C}
{Carrasco}, J.~M., {Weiler}, M., {Jordi}, C., {et~al.} 2021, \aap, 652, A86

\bibitem[{{Chojnowski} {et~al.}(2019){Chojnowski}, {Hubrig}, {Hasselquist},
  {Castelli}, {Whelan}, {Majewski}, {Nitschelm}, {Garc{\'\i}a-Hern{\'a}ndez},
  {Stassun}, \& {Zamora}}]{2019ApJ...873L...5C}
{Chojnowski}, S.~D., {Hubrig}, S., {Hasselquist}, S., {et~al.} 2019, \apjl,
  873, L5

\bibitem[{{Cui} {et~al.}(2012){Cui}, {Zhao}, {Chu}, {Li}, {Li}, {Zhang}, {Su},
  {Yao}, {Wang}, {Xing}, {Li}, {Zhu}, {Wang}, {Gu}, {Luo}, {Xu}, {Zhang},
  {Liu}, {Zhang}, {Yang}, {Cao}, {Chen}, {Chen}, {Chen}, {Chen}, {Chu}, {Feng},
  {Gong}, {Hou}, {Hu}, {Hu}, {Hu}, {Jia}, {Jiang}, {Jiang}, {Jiang}, {Jin},
  {Li}, {Li}, {Li}, {Liu}, {Liu}, {Lu}, {Mao}, {Men}, {Qi}, {Qi}, {Shi},
  {Tang}, {Tao}, {Wang}, {Wang}, {Wang}, {Wang}, {Wang}, {Wang}, {Wang},
  {Wang}, {Wang}, {Wang}, {Wang}, {Wang}, {Xu}, {Xu}, {Yang}, {Yu}, {Yuan},
  {Yuan}, {Zhai}, {Zhang}, {Zhang}, {Zhang}, {Zhao}, {Zhou}, {Zhou}, {Zhu}, \&
  {Zou}}]{2012RAA....12.1197C}
{Cui}, X.-Q., {Zhao}, Y.-H., {Chu}, Y.-Q., {et~al.} 2012, Research in Astronomy
  and Astrophysics, 12, 1197

\bibitem[{{Gaia Collaboration} {et~al.}(2022){Gaia Collaboration}, {Vallenari},
  {Brown}, {Prusti}, {de Bruijne}, {Arenou}, {Babusiaux}, {Biermann},
  {Creevey}, {Ducourant}, {Evans}, {Eyer}, {Guerra}, {Hutton}, {Jordi},
  {Klioner}, {Lammers}, {Lindegren}, {Luri}, {Mignard}, {Panem}, {Pourbaix},
  {Randich}, {Sartoretti}, {Soubiran}, {Tanga}, {Walton}, {Bailer-Jones},
  {Bastian}, {Drimmel}, {Jansen}, {Katz}, {Lattanzi}, {van Leeuwen}, {Bakker},
  {Cacciari}, {Casta{\~n}eda}, {De Angeli}, {Fabricius}, {Fouesneau},
  {Fr{\'e}mat}, {Galluccio}, {Guerrier}, {Heiter}, {Masana}, {Messineo},
  {Mowlavi}, {Nicolas}, {Nienartowicz}, {Pailler}, {Panuzzo}, {Riclet}, {Roux},
  {Seabroke}, {Sordo{\o}rcit}, {Th{\'e}venin}, {Gracia-Abril}, {Portell},
  {Teyssier}, {Altmann}, {Andrae}, {Audard}, {Bellas-Velidis}, {Benson},
  {Berthier}, {Blomme}, {Burgess}, {Busonero}, {Busso}, {C{\'a}novas}, {Carry},
  {Cellino}, {Cheek}, {Clementini}, {Damerdji}, {Davidson}, {de Teodoro},
  {Nu{\~n}ez Campos}, {Delchambre}, {Dell'Oro}, {Esquej},
  {Fern{\'a}ndez-Hern{\'a}ndez}, {Fraile}, {Garabato}, {Garc{\'\i}a-Lario},
  {Gosset}, {Haigron}, {Halbwachs}, {Hambly}, {Harrison}, {Hern{\'a}ndez},
  {Hestroffer}, {Hodgkin}, {Holl}, {Jan{\ss}en}, {Jevardat de Fombelle},
  {Jordan}, {Krone-Martins}, {Lanzafame}, {L{\"o}ffler}, {Marchal}, {Marrese},
  {Moitinho}, {Muinonen}, {Osborne}, {Pancino}, {Pauwels}, {Recio-Blanco},
  {Reyl{\'e}}, {Riello}, {Rimoldini}, {Roegiers}, {Rybizki}, {Sarro}, {Siopis},
  {Smith}, {Sozzetti}, {Utrilla}, {van Leeuwen}, {Abbas}, {{\'A}brah{\'a}m},
  {Abreu Aramburu}, {Aerts}, {Aguado}, {Ajaj}, {Aldea-Montero}, {Altavilla},
  {{\'A}lvarez}, {Alves}, {Anders}, {Anderson}, {Anglada Varela}, {Antoja},
  {Baines}, {Baker}, {Balaguer-N{\'u}{\~n}ez}, {Balbinot}, {Balog}, {Barache},
  {Barbato}, {Barros}, {Barstow}, {Bartolom{\'e}}, {Bassilana}, {Bauchet},
  {Becciani}, {Bellazzini}, {Berihuete}, {Bernet}, {Bertone}, {Bianchi},
  {Binnenfeld}, {Blanco-Cuaresma}, {Blazere}, {Boch}, {Bombrun}, {Bossini},
  {Bouquillon}, {Bragaglia}, {Bramante}, {Breedt}, {Bressan}, {Brouillet},
  {Brugaletta}, {Bucciarelli}, {Burlacu}, {Butkevich}, {Buzzi}, {Caffau},
  {Cancelliere}, {Cantat-Gaudin}, {Carballo}, {Carlucci}, {Carnerero},
  {Carrasco}, {Casamiquela}, {Castellani}, {Castro-Ginard}, {Chaoul},
  {Charlot}, {Chemin}, {Chiaramida}, {Chiavassa}, {Chornay}, {Comoretto},
  {Contursi}, {Cooper}, {Cornez}, {Cowell}, {Crifo}, {Cropper}, {Crosta},
  {Crowley}, {Dafonte}, {Dapergolas}, {David}, {David}, {de Laverny}, {De
  Luise}, {De March}, {De Ridder}, {de Souza}, {de Torres}, {del Peloso}, {del
  Pozo}, {Delbo}, {Delgado}, {Delisle}, {Demouchy}, {Dharmawardena}, {Di
  Matteo}, {Diakite}, {Diener}, {Distefano}, {Dolding}, {Edvardsson}, {Enke},
  {Fabre}, {Fabrizio}, {Faigler}, {Fedorets}, {Fernique}, {Fienga}, {Figueras},
  {Fournier}, {Fouron}, {Fragkoudi}, {Gai}, {Garcia-Gutierrez},
  {Garcia-Reinaldos}, {Garc{\'\i}a-Torres}, {Garofalo}, {Gavel}, {Gavras},
  {Gerlach}, {Geyer}, {Giacobbe}, {Gilmore}, {Girona}, {Giuffrida}, {Gomel},
  {Gomez}, {Gonz{\'a}lez-N{\'u}{\~n}ez}, {Gonz{\'a}lez-Santamar{\'\i}a},
  {Gonz{\'a}lez-Vidal}, {Granvik}, {Guillout}, {Guiraud},
  {Guti{\'e}rrez-S{\'a}nchez}, {Guy}, {Hatzidimitriou}, {Hauser}, {Haywood},
  {Helmer}, {Helmi}, {Sarmiento}, {Hidalgo}, {Hilger}, {H{\l}adczuk}, {Hobbs},
  {Holland}, {Huckle}, {Jardine}, {Jasniewicz}, {Jean-Antoine Piccolo},
  {Jim{\'e}nez-Arranz}, {Jorissen}, {Juaristi Campillo}, {Julbe}, {Karbevska},
  {Kervella}, {Khanna}, {Kontizas}, {Kordopatis}, {Korn}, {K{\'o}sp{\'a}l},
  {Kostrzewa-Rutkowska}, {Kruszy{\'n}ska}, {Kun}, {Laizeau}, {Lambert},
  {Lanza}, {Lasne}, {Le Campion}, {Lebreton}, {Lebzelter}, {Leccia}, {Leclerc},
  {Lecoeur-Taibi}, {Liao}, {Licata}, {Lindstr{\o}m}, {Lister}, {Livanou},
  {Lobel}, {Lorca}, {Loup}, {Madrero Pardo}, {Magdaleno Romeo}, {Managau},
  {Mann}, {Manteiga}, {Marchant}, {Marconi}, {Marcos}, {Marcos Santos},
  {Mar{\'\i}n Pina}, {Marinoni}, {Marocco}, {Marshall}, {Polo},
  {Mart{\'\i}n-Fleitas}, {Marton}, {Mary}, {Masip}, {Massari},
  {Mastrobuono-Battisti}, {Mazeh}, {McMillan}, {Messina}, {Michalik}, {Millar},
  {Mints}, {Molina}, {Molinaro}, {Moln{\'a}r}, {Monari}, {Mongui{\'o}},
  {Montegriffo}, {Montero}, {Mor}, {Mora}, {Morbidelli}, {Morel}, {Morris},
  {Muraveva}, {Murphy}, {Musella}, {Nagy}, {Noval}, {Oca{\~n}a}, {Ogden},
  {Ordenovic}, {Osinde}, {Pagani}, {Pagano}, {Palaversa}, {Palicio},
  {Pallas-Quintela}, {Panahi}, {Payne-Wardenaar}, {Pe{\~n}alosa Esteller},
  {Penttil{\"a}}, {Pichon}, {Piersimoni}, {Pineau}, {Plachy}, {Plum}, {Poggio},
  {Pr{\v{s}}a}, {Pulone}, {Racero}, {Ragaini}, {Rainer}, {Raiteri}, {Rambaux},
  {Ramos}, {Ramos-Lerate}, {Re Fiorentin}, {Regibo}, {Richards}, {Rios Diaz},
  {Ripepi}, {Riva}, {Rix}, {Rixon}, {Robichon}, {Robin}, {Robin}, {Roelens},
  {Rogues}, {Rohrbasser}, {Romero-G{\'o}mez}, {Rowell}, {Royer}, {Ruz Mieres},
  {Rybicki}, {Sadowski}, {S{\'a}ez N{\'u}{\~n}ez}, {Sagrist{\`a} Sell{\'e}s},
  {Sahlmann}, {Salguero}, {Samaras}, {Sanchez Gimenez}, {Sanna},
  {Santove{\~n}a}, {Sarasso}, {Schultheis}, {Sciacca}, {Segol}, {Segovia},
  {S{\'e}gransan}, {Semeux}, {Shahaf}, {Siddiqui}, {Siebert}, {Siltala},
  {Silvelo}, {Slezak}, {Slezak}, {Smart}, {Snaith}, {Solano}, {Solitro},
  {Souami}, {Souchay}, {Spagna}, {Spina}, {Spoto}, {Steele},
  {Steidelm{\"u}ller}, {Stephenson}, {S{\"u}veges}, {Surdej}, {Szabados},
  {Szegedi-Elek}, {Taris}, {Taylo}, {Teixeira}, {Tolomei}, {Tonello}, {Torra},
  {Torra}, {Torralba Elipe}, {Trabucchi}, {Tsounis}, {Turon}, {Ulla}, {Unger},
  {Vaillant}, {van Dillen}, {van Reeven}, {Vanel}, {Vecchiato}, {Viala},
  {Vicente}, {Voutsinas}, {Weiler}, {Wevers}, {Wyrzykowski}, {Yoldas}, {Yvard},
  {Zhao}, {Zorec}, {Zucker}, \& {Zwitter}}]{2022arXiv220800211G}
{Gaia Collaboration}, {Vallenari}, A., {Brown}, A.~G.~A., {et~al.} 2022, arXiv
  e-prints, arXiv:2208.00211

\bibitem[{{Green} {et~al.}(2019){Green}, {Schlafly}, {Zucker}, {Speagle}, \&
  {Finkbeiner}}]{2019ApJ...887...93G}
{Green}, G.~M., {Schlafly}, E., {Zucker}, C., {Speagle}, J.~S., \&
  {Finkbeiner}, D. 2019, \apj, 887, 93

\bibitem[{{Holdsworth}(2021)}]{2021mobs.confE..27H}
{Holdsworth}, D.~L. 2021, in MOBSTER-1 virtual conference: Stellar Variability
  as a Probe of Magnetic Fields in Massive Stars, 27

\bibitem[{{H{\"u}mmerich} {et~al.}(2020){H{\"u}mmerich}, {Paunzen}, \&
  {Bernhard}}]{2020A&A...640A..40H}
{H{\"u}mmerich}, S., {Paunzen}, E., \& {Bernhard}, K. 2020, \aap, 640, A40

\bibitem[{{Khan} \& {Shulyak}(2006)}]{2006A&A...448.1153K}
{Khan}, S.~A. \& {Shulyak}, D.~V. 2006, \aap, 448, 1153

\bibitem[{{Khan} \& {Shulyak}(2007)}]{2007A&A...469.1083K}
{Khan}, S.~A. \& {Shulyak}, D.~V. 2007, \aap, 469, 1083

\bibitem[{{Kudryavtsev} {et~al.}(2006){Kudryavtsev}, {Romanyuk}, {Elkin}, \&
  {Paunzen}}]{2006MNRAS.372.1804K}
{Kudryavtsev}, D.~O., {Romanyuk}, I.~I., {Elkin}, V.~G., \& {Paunzen}, E. 2006,
  \mnras, 372, 1804

\bibitem[{{Kupka} {et~al.}(2004){Kupka}, {Paunzen}, {Iliev}, \&
  {Maitzen}}]{2004MNRAS.352..863K}
{Kupka}, F., {Paunzen}, E., {Iliev}, I.~K., \& {Maitzen}, H.~M. 2004, \mnras,
  352, 863

\bibitem[{{Maitzen} \& {Seggewiss}(1980)}]{1980A&A....83..328M}
{Maitzen}, H.~M. \& {Seggewiss}, W. 1980, \aap, 83, 328

\bibitem[{{Montegriffo} {et~al.}(2022){Montegriffo}, {De Angeli}, {Andrae},
  {Riello}, {Pancino}, {Sanna}, {Bellazzini}, {Evans}, {Carrasco}, {Sordo},
  {Busso}, {Cacciari}, {Jordi}, {van Leeuwen}, {Vallenari}, {Altavilla},
  {Barstow}, {Brown}, {Burgess}, {Castellani}, {Cowell}, {Davidson}, {De
  Luise}, {Delchambre}, {Diener}, {Fabricius}, {Fremat}, {Fouesneau},
  {Gilmore}, {Giuffrida}, {Hambly}, {Harrison}, {Hidalgo}, {Hodgkin},
  {Holland}, {Marinoni}, {Osborne}, {Pagani}, {Palaversa}, {Piersimoni},
  {Pulone}, {Ragaini}, {Rainer}, {Richards}, {Rowell}, {Ruz-Mieres}, {Sarro},
  {Walton}, \& {Yoldas}}]{2022arXiv220606205M}
{Montegriffo}, P., {De Angeli}, F., {Andrae}, R., {et~al.} 2022, arXiv
  e-prints, arXiv:2206.06205

\bibitem[{{Netopil} {et~al.}(2008){Netopil}, {Paunzen}, {Maitzen}, {North}, \&
  {Hubrig}}]{2008A&A...491..545N}
{Netopil}, M., {Paunzen}, E., {Maitzen}, H.~M., {North}, P., \& {Hubrig}, S.
  2008, \aap, 491, 545

\bibitem[{{Netopil} {et~al.}(2007){Netopil}, {Paunzen}, {Maitzen}, {Pintado},
  {Claret}, {Miranda}, {Iliev}, \& {Casanova}}]{2007A&A...462..591N}
{Netopil}, M., {Paunzen}, E., {Maitzen}, H.~M., {et~al.} 2007, \aap, 462, 591

\bibitem[{{Paunzen}(2020)}]{2020CoSka..50..570P}
{Paunzen}, E. 2020, Contributions of the Astronomical Observatory Skalnate
  Pleso, 50, 570

\bibitem[{{Paunzen} {et~al.}(2014){Paunzen}, {Netopil}, {Maitzen}, {Pavlovski},
  {Schnell}, \& {Zejda}}]{2014A&A...564A..42P}
{Paunzen}, E., {Netopil}, M., {Maitzen}, H.~M., {et~al.} 2014, \aap, 564, A42

\bibitem[{{Paunzen} {et~al.}(2005{\natexlab{a}}){Paunzen}, {Pintado},
  {Maitzen}, \& {Claret}}]{2005MNRAS.362.1025P}
{Paunzen}, E., {Pintado}, O.~I., {Maitzen}, H.~M., \& {Claret}, A.
  2005{\natexlab{a}}, \mnras, 362, 1025

\bibitem[{{Paunzen} {et~al.}(2005{\natexlab{b}}){Paunzen}, {Schnell}, \&
  {Maitzen}}]{2005A&A...444..941P}
{Paunzen}, E., {Schnell}, A., \& {Maitzen}, H.~M. 2005{\natexlab{b}}, \aap,
  444, 941

\bibitem[{{Paunzen} {et~al.}(2006){Paunzen}, {Schnell}, \&
  {Maitzen}}]{2006A&A...458..293P}
{Paunzen}, E., {Schnell}, A., \& {Maitzen}, H.~M. 2006, \aap, 458, 293

\bibitem[{{Paunzen} {et~al.}(2005{\natexlab{c}}){Paunzen}, {St{\"u}tz}, \&
  {Maitzen}}]{2005A&A...441..631P}
{Paunzen}, E., {St{\"u}tz}, C., \& {Maitzen}, H.~M. 2005{\natexlab{c}}, \aap,
  441, 631

\bibitem[{{Preston}(1974)}]{1974ARA&A..12..257P}
{Preston}, G.~W. 1974, \araa, 12, 257

\bibitem[{{Richer} {et~al.}(2000){Richer}, {Michaud}, \&
  {Turcotte}}]{2000ApJ...529..338R}
{Richer}, J., {Michaud}, G., \& {Turcotte}, S. 2000, \apj, 529, 338

\bibitem[{{Romanyuk} \& {Kudryavtsev}(2008)}]{2008AstBu..63..139R}
{Romanyuk}, I.~I. \& {Kudryavtsev}, D.~O. 2008, Astrophysical Bulletin, 63, 139

\bibitem[{{Stigler} {et~al.}(2014){Stigler}, {Maitzen}, {Paunzen}, \&
  {Netopil}}]{2014A&A...562A..65S}
{Stigler}, C., {Maitzen}, H.~M., {Paunzen}, E., \& {Netopil}, M. 2014, \aap,
  562, A65

\bibitem[{{Th{\'e}ado} {et~al.}(2005){Th{\'e}ado}, {Vauclair}, \&
  {Cunha}}]{2005A&A...443..627T}
{Th{\'e}ado}, S., {Vauclair}, S., \& {Cunha}, M.~S. 2005, \aap, 443, 627

\bibitem[{{Zavada} \& {P{\'\i}{\v{s}}ka}(2022)}]{2022AJ....163...33Z}
{Zavada}, P. \& {P{\'\i}{\v{s}}ka}, K. 2022, \aj, 163, 33

\end{thebibliography}

\end{document}